# Compact bilateral single conductor surface wave transmission line

Zhixia Xu, Shunli Li, Hongxin Zhao, Leilei Liu and Xiaoxing Yin

A compact bilateral single conductor surface wave transmission line (TL) is proposed, converting the quasi-transverse electromagnetic (QTEM) mode of low characteristic impedance slotline into the transverse magnetic (TM) mode of single-conductor TL. The propagation constant of the proposed TL is decided by geometric parameters of the periodic corrugated structure. Compared to conventional transitions between coplanar waveguide (CPW) and single-conductor TLs, such as Goubau line (G-Line) and surface plasmons TL, the proposed structure halves the size and this feature gives important potentials to develop integrated surface wave devices and circuits. The designed structure, sample fabrication and experimental verification are discussed.

*Introduction:* The area of surface wave transmission lines (TL) has been developed for many years, and how to excite and transmit surface waves are still hot topics recently [1-3]. Single-conductor TLs, as a kind of open waveguide, are widely used to support surface waves. The well-known Goubau line (G-Line) was proposed by Goubau and featured by less bulk, less expensive and smaller dielectric and conductor loss than closed guide [4]. Another significant milestone is the proposal of the spoof surface plasmon polaritons (SSPPs) whose dispersion characteristic could be carefully designed by geometric dimensions of periodic structure. We can confine electronic magnetic (EM) field and tailor dispersion curves arbitrarily by designing geometric dimensions of periodic structure accordingly [5, 6]. To some extent, G-Line and SSPPs TL are both a kind of single conductor surface wave TL.

Conventional single-conductor TLs consist of a metallic strip and a pair of mode transitions. Using a pair of flaring ground planes, surface modes can be excited [1]. But the large size of the preceding transitions between coplanar waveguide (CPW) and single-conductor guide is always a drawback to integrated circuits. In this letter, a novel transition between bilateral slotline [7] and bilateral corrugated metallic strip is proposed to reduce the size of transition to half the size of preceding designs. The bilateral copper structure is fabricated on Rogers 4003, and measured results agree well with simulated results well.

*Structure design:* The proposed bilateral TL consists of three parts, low characteristic impedance slotline, a pair of transitions and a corrugated metallic strip. The whole schematic configuration is shown in Fig. 1. The total length $L$ is 175 mm; the total width $W$ is 20.3 mm; the width of the flaring ground $Wg$ is 15 mm. The thickness of copper layer is 0.035 mm, and same structures at top and bottom sides are connected by via arrays, whose radius are 0.2mm, and printed on a 1.524 mm thick Rogers 4003 substrate.

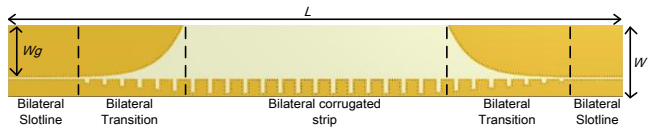

**Fig. 1** *Configuration of the compact bilateral surface wave TL*

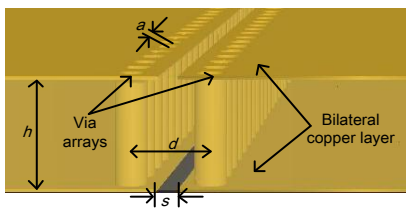

**Fig. 2** *Configuration of the low characteristic impedance slotline*

Regarded as half mode CPW, a series of slotlines have been proposed whose characteristic impedance is easy to control [7]. The bilateral slotline with 50 ohm characteristic impedance is designed as shown in Fig. 2. The copper layers on both sides of the substrate are connected by the metalized via arrays, and the two via arrays can increase the perunit-length capacitance and decrease characteristic impedance to 50 ohm which is usually quite high for a conventional slotline. The height of via $h$ is 1.524 mm, the slot $S$ is 0.3 mm; the space between via arrays $d$ is 0.4 mm; the space between neighbour vias $a$ is 1 mm.

As a single conductor TL, bilateral corrugated metallic strips are printed on both sides of the substrate and connected by via arrays the configuration is shown in Fig. 3a. To excite the transverse magnetic (TM) mode, the surface mode of the corrugated strip, from the quasi-transverse electromagnetic (QTEM) mode, the mode of slotline, we propose a compact bilateral transition structure, as shown in Fig. 3b. The detail dimensions and parameters of the proposed bilateral surface wave TL are given in Table 1.

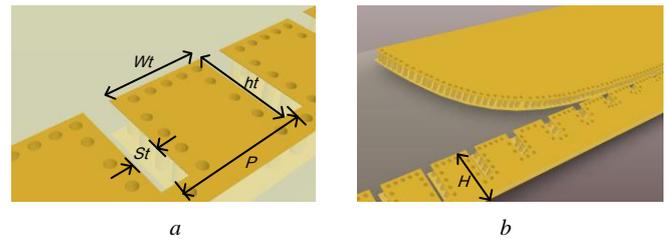

**Fig. 3** *The way to excite the surface mode*
a  Bilateral corrugated metallic strip
b  Compact bilateral transition

**Table 1:** Dimensions of the proposed transition (units: millimetres)

| L | W | Wg | h | a | d |
|---|---|----|---|---|---|
| 175 | 20.3 | 15 | 1.524 | 1 | 0.4 |
| s | H | Wt | Ht | St | P |
| 0.3 | 5 | 4 | 4 | 1 | 5 |

*Results and discussion:* The configuration of the sample is shown in Fig. 4. The bilateral layer structure is 0.035 mm thick copper printed on both sides of a 1.524mm thick Rogers 4003 substrate with relative permittivity of 3.55. Two SMA connectors are welded on both terminals of the fabricated sample, and two 50 Ohm coaxial lines are utilized to connect the sample with Vector Network Analyser to measure S-parameters.

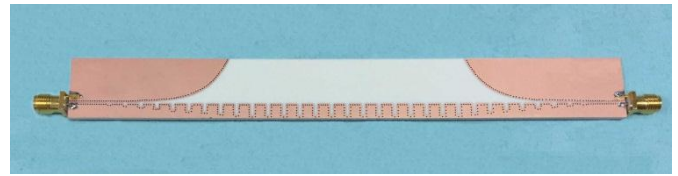

**Fig. 4** *Fabrication sample*

Dispersion curve is calculated by CST Microwave Studio, and shown in Fig. 5a. The curve lies outside the light cone, indicating that wave propagating along the TL is a kind of slow wave. Fig. 5b and Fig. 5c compare simulated S-parameters curves and measured S-parameters curves respectively. The transmission loss mainly comes from radiation loss. Both simulated and measured curves show that reflection coefficient $S_{11}$ is below −10dB. The transmission coefficient $S_{21}$ is above −7dB at the full passing band, and increases to −3dB around 10GHz the cutoff frequency. Because as the frequency rises, the propagation constant becomes larger and the surface mode is bounded more tightly around the corrugated strip that means the radiation loss from the surface mode decreases, so the proposed TL can transmit EM waves efficiently from 7GHz to 10GHz.

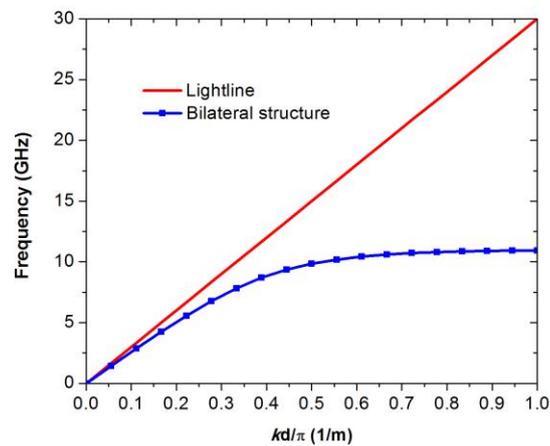

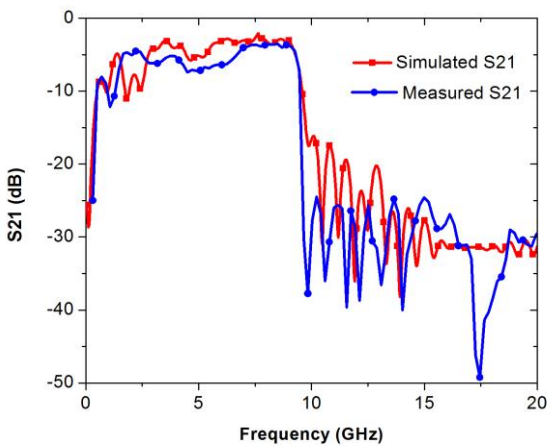

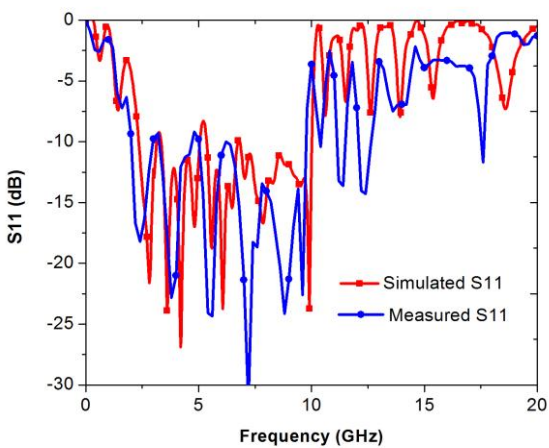

**Fig. 5** *Fabrication sample and its S-parameters*
*a* Dispersion curve
*b* S21
*c* S11

*Conclusion:* This paper presents a compact bilateral single conductor surface wave TL, halving the size of conventional designs. The compact structure has potentials in the high integrated surface wave devices. Nevertheless, the flaring transition of the proposed structure can be regarded as a kind of tapered slot antenna (TSA), and radiation effect is more considerable than preceding designs based on CPWs which can be regarded as a pair of TSAs with differential beam whose radiation efficiency is much less. Research on suppression radiation loss from this compact surface wave TL remains to be done.

*Acknowledgments:* This work was supported in part by National Natural Science Foundation of China under Grant 61427801, in part by National Natural Science Foundation of China under Grant U1536123, in part by National Natural Science Foundation of China under Grant U1536124.

Zhixia Xu, Shunli Li, Hongxin Zhao and Xiaoxing Yin (State Key Laboratory of Millimeter Waves, Southeast University, Nanjing 210096, People's Republic of China)
E-mail: 101010074@seu.edu.cn

Leilei Liu (School of Electronic Science and Engineering, Nanjing University of Posts and Telecommunications, Nanjing 210003, People's Republic of China)